\documentstyle[eqsecnum,preprint,aps,prd,psfig]{revtex}

\begin{document}


\def\bq{\begin{equation}}
\def\nq{\end{equation}}
\def\bqr{\begin{eqnarray}}
\def\nqr{\end{eqnarray}}
\let\l=\left
\let\r=\right


\title{Spacetime structure of an inflating global monopole}
\author{
Inyong Cho \footnote [1] {Electronic address: cho@cosmos2.phy.tufts.edu} 
and Alexander Vilenkin \footnote [2] 
{Electronic address: vilenkin@cosmos2.phy.tufts.edu} }

\address{Tufts Institute of Cosmology\\
	Department of Physics and Astronomy\\
	Tufts University\\
	Medford, MA 02155}
\date{\today}
\maketitle

\begin{abstract}

The evolution of a global monopole with an inflating core is investigated.
An analytic expression for the exterior metric at large distances
from the core is obtained.
The overall spacetime structure is studied numerically, both in vacuum 
and in a radiation background.

\end{abstract}

\section{Introduction}

It has been argued in ref. \cite{AL,AV} that inflation can occur in the
cores of topological defects. The condition for this is that the symmetry 
breaking scale of the defects satisfy $\eta > \eta_c \sim {\cal O} (m_p)$.
An attractive feature of this topological inflation is that it does not
require fine-tuning of the initial conditions. The qualitative arguments of 
refs. \cite{AL,AV} were later verified in numerical simulations by
Sakai {\it et. al.} \cite{NS}.
They found in particular that the critical value of $\eta$ for
domain walls and global monopoles is $\eta_c \simeq 0.33\,m_p$.

In this paper, we would like to discuss the spacetime structure of 
inflating defects. This issue was partially addressed in refs. \cite{AV,NS}.
However, some questions still remain unanswered. In particular, it is not
clear what inflating defects look like from the outside.
To be specific, consider the case of global monopoles.
For $\eta < \eta_c$, the metric at large distances from the core has 
a solid deficit angle $\Delta \simeq 4\pi(8\pi G \eta^2)$
\cite{BV}. When $\eta > \eta_c$, the solid deficit angle exceeds $4\pi$
and static solutions do not exist.
It was conjectured in ref. \cite{AV} and verified in ref. \cite{NS} that
static solutions cease to exist at the same value of $\eta = \eta_c$
for which topological inflation becomes possible.
The problem is then to determine the metric outside the inflating monopole.
We shall address this problem both analytically and numerically.

In the next section, we review the static monopole solution of ref. \cite{BV}
and conjecture that the metric we are looking for is obtained by
continuing this static metric to $\Delta > 4\pi$.
This conjecture is then verified in section III by numerically 
solving combined Einstein's and scalar field equations. 
(Here, we follow the technique of Sakai {\it et. al.})
The overall spacetime structure of the inflating monopole is discussed
in section IV.
In section V, we study the global monopole dynamics with cosmological
initial condtions which include a radiation background. Our conclusions are
summarized in section VI.

\section{Asymptotic metric}

The simplest model that gives rise to global monopoles is described by 
the Lagrangian

\bq
	{\cal L} = -{1\over 2}\partial_\mu\phi^a\partial^\mu\phi^a -
		{1 \over 4} \lambda (\phi^a\phi^a - \eta^2)^2\,,
\nq
where $\phi^a$ is a triplet of scalar fields, $a = 1,2,3$.
The model has a global O(3) symmetry spontaneously broken down to U(1).
A global monopole of unit topological charge 
is described by the ``hedgehog'' configuration
$\phi^a = \phi (r)\, \hat x^a$, where $\hat x^a$ is a radial 
unit vector.  Outside the monopole core, $\phi (r) \approx \eta$.
By soving Einstein's eq. in the asymptotic region outside the core, 
the metric is found to be \cite{BV}

\bq
	ds^2 = - \l( 1 - 8\pi G\eta^2 - {2GM \over R} \r)\,dT^2
		+ \l( 1 - 8\pi G\eta^2 - {2GM \over R} \r)^{-1}\,dR^2
		+R^2 d\Omega^2\,.
\label{eq=manuelmet}
\nq
(We use $\hbar = c = 1\,,G = 1/m_p^2\,.$)

At large distances from the core, the mass term can be neglected, and after 
rescaling $T$ and $R$ coordinates the metric takes the form

\bq
	ds^2 = - dT'^2 + dR'^2 + (1 - 8\pi G\eta^2)\,R'^2 d\Omega^2\,.
\nq
This metric exhibits a solid angle deficit
$\Delta = 4\pi (8\pi G\eta^2)$.
Another useful form of the global monopole metric can be obtained
by a coordinate transformation

\bq
	t = {1 \over \sqrt{|1-v^2|}} (T' - vR')\,, \qquad 
	r = {1 \over \sqrt{|1-v^2|}} (R' - vT')\,,
\label{eq=cdtransf}
\nq
where, $v = \sqrt{8\pi G\eta^2}$. This gives

\bq
	ds^2 = -dt^2 + dr^2 + (r + \sqrt{8\pi G\eta^2}\, t)^2 d\Omega^2\,.
\label{eq=asympmet}
\nq

Let us now formally consider the metric (\ref{eq=manuelmet}) with 
a solid deficit angle $\Delta > 4\pi$. In this case,
the asymptotic form of the metric is

\bq
	ds^2 = - dR'^2 + dT'^2 + (8\pi G\eta^2 - 1)\,R'^2 d\Omega^2\,.
\label{eq=primemet}
\nq
Here, $T'$ is a spacelike and $R'$ is a timelike coordinate.
The natural ranges of these coordinates are
$-\infty < T' < \infty$ and $0 < R' < \infty$.
The metric (\ref{eq=primemet}) then represents a ``cylindrical'' universe 
of topology $R \times S(2)$. 
The expansion of this universe is highly anisotropic:
there is no expansion in the $T'$ direction along the axis of the cylinder,
while the radius of the spherical sections grows proportionally to
time $(R')$.

The coordinate transformation (\ref{eq=cdtransf}) brings (\ref{eq=primemet}) 
to the form (\ref{eq=asympmet}). 
Note that the metric (\ref{eq=asympmet}) applies to both $\Delta < 4\pi$ and 
$\Delta > 4\pi$. It is easily verified that (\ref{eq=asympmet}) 
solves Einstein's equations
with energy-momentum tensor corresponding to the ansatz 
$\phi^a = \phi (r)\, \hat x^a$.

It seems reasonable to assume that equations (\ref{eq=asympmet}), 
(\ref{eq=primemet}) represent the 
exterior asymptotic region 
of an inflating monopole. In the next section, we shall verify this 
assumption by numerically solving Einstein's and scalar field equations.

\section{Numerical results}

In this section, we use the technique of Sakai {\it et. al.}\cite{NS} 
to study the monopole evolution numerically. 
We use the general spherically-symmetric ansatz for the metric 

\bq
	ds^2 = - dt^2 + A(t,r)^2\,dr^2 +B(t,r)^2\,r^2\,d\Omega^2\,,
\label{eq=genmet}
\nq
and the scalar field,

\bq
	\phi^a = \phi (t,r)\, \hat x^a	\,.
\nq
The corresponding field eqs. are shown in Appendix.

Following Sakai {\it et. al.}, we set up the initial conditions by
assuming the 3-metric to be flat at
the initial moment, $t = 0$.  The initial monopole field $\phi(0,r)$
is obtained by numerically solving the static flat-space field
equation with the boundary conditions $\phi(r=0) = 0$
and $\phi(r=\infty) = \eta$.  Finally,we set ${\dot \phi (0,r)}=0$ and
evaluate ${\dot A(0,r)}$ and ${\dot B(0,r)}$ from the Hamiltonian and the
momentum constraints (\ref{eq=energy}),(\ref{eq=K2}).

We solved the field equations with these initial conditions for
several values of the symmetry breaking scale $\eta$.  Our results are
in full agreement with those of Sakai {\it et. al.} \cite{NS}.
While the latter authors concentrated mainly on the inflating region
in the monopole core, we shall analyze the asymptotic region and the
overall structure of the monopole spacetime.

\begin{figure}
\psfig{file=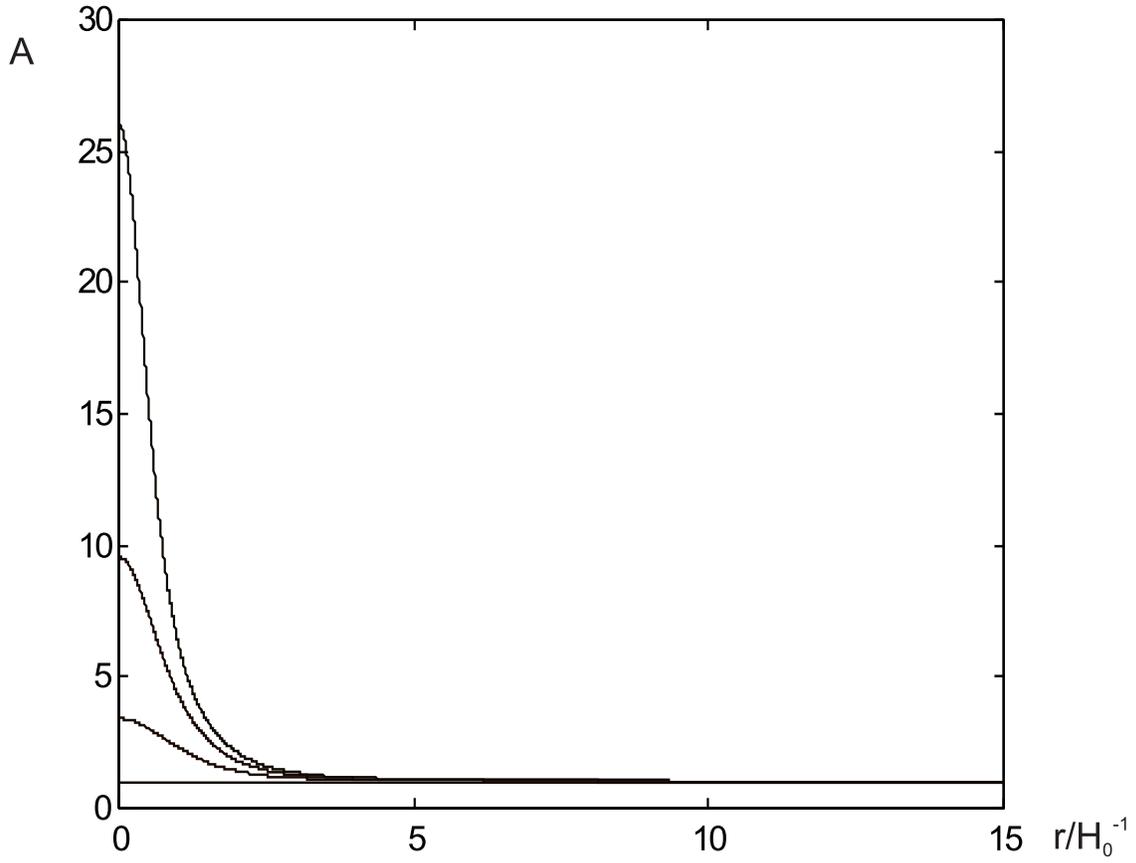}
\caption{
A plot of $A(t,r)$ vs. $H_0r$ at $H_0t=1,2,3$ for $\eta = 0.6\,m_p$.
Lower curves correspond to earlier times.
At small $r$, $A(t,r)$ grows rapidly with time.
In the asymptotic region, $A(t,r) \approx 1$.
}
\label{fig=mona}
\end{figure}

\begin{figure}
\psfig{file=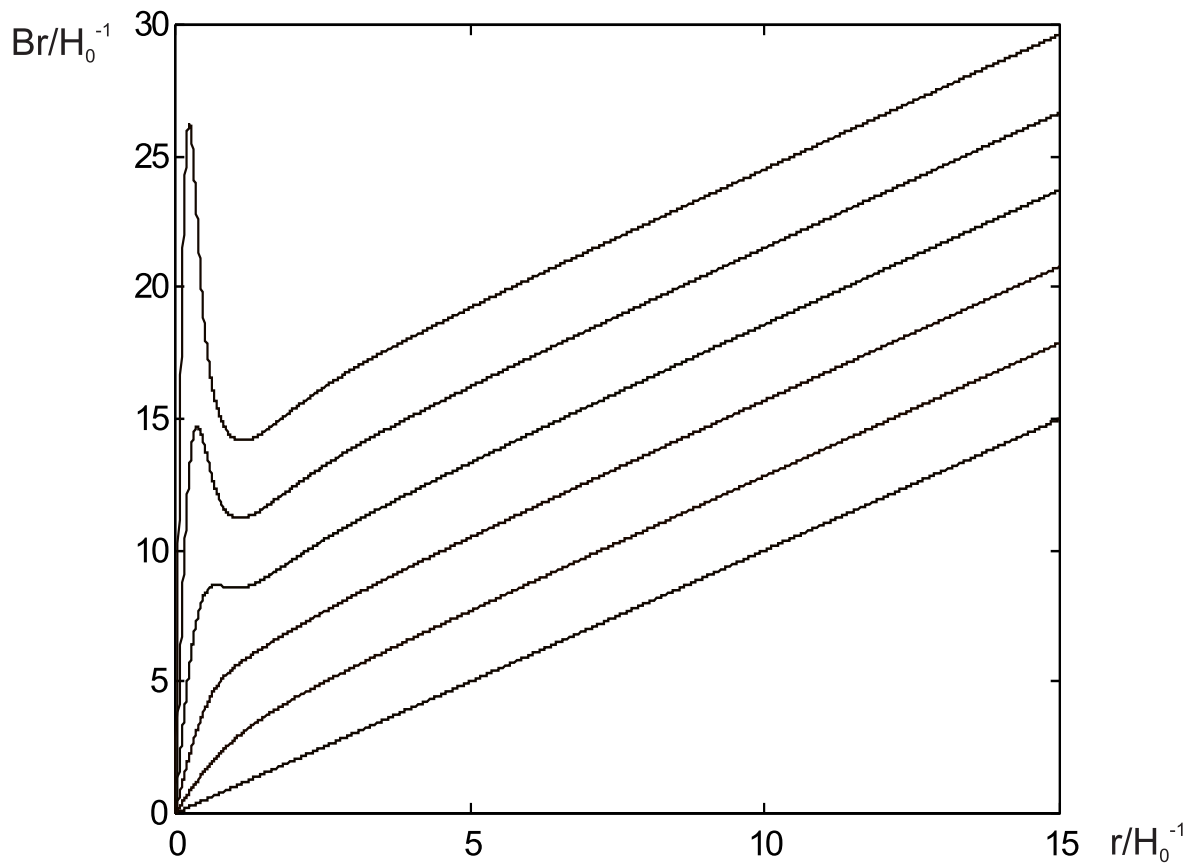}
\caption{
A plot of $H_0rB(t,r)$ vs. $H_0r$ at $H_0t=0,1,2,3,4,5$ for $\eta = 0.6\,m_p$.
Lower curves correspond to earlier times.
In the asymptotic region, lines are equally spaced and the separation is
$\approx \sqrt{8\pi G\eta^2}$.
}
\label{fig=monb}
\end{figure}

To illustrate our results, we shall take $\eta=0.6\, m_p$, which is
greater than the critical value $\eta_c=0.33\, m_p$.  The solutions
$A(t,r)$ and $B(t,r)\,r$ are shown as functions of $r$ at several moments 
of time in figures \ref{fig=mona} and \ref{fig=monb}, respectively.
In the figures, $r$ and $t$ are shown in units of $H_0^{-1} =(8\pi
G V(\phi=0)/3)^{1/2}$, which is the horizon radius at the monopole center.
At small $r$ we clearly have inflation: both
$A$ and $B$ rapidly grow with time, with $A\approx B$.  

Now, it is easily verified that for $r\gtrsim 4H_0^{-1}$ the metric is
well approximated by Eq.(\ref{eq=asympmet}): 
Fig.\ref{fig=mona} shows that $A(t,r) \approx 1$ , and Fig.\ref{fig=monb} 
shows that the graphs of $B(t,r)\,r$ at $H_0t=0,1,2,3,4,5$ are
equally spaced straight lines.  The quantitative agreement with the
coefficient $\sqrt{8\pi G\eta^2}$ in (\ref{eq=asympmet}) 
is also easily checked.

\section{Spacetime structure}

\begin{figure}
\psfig{file=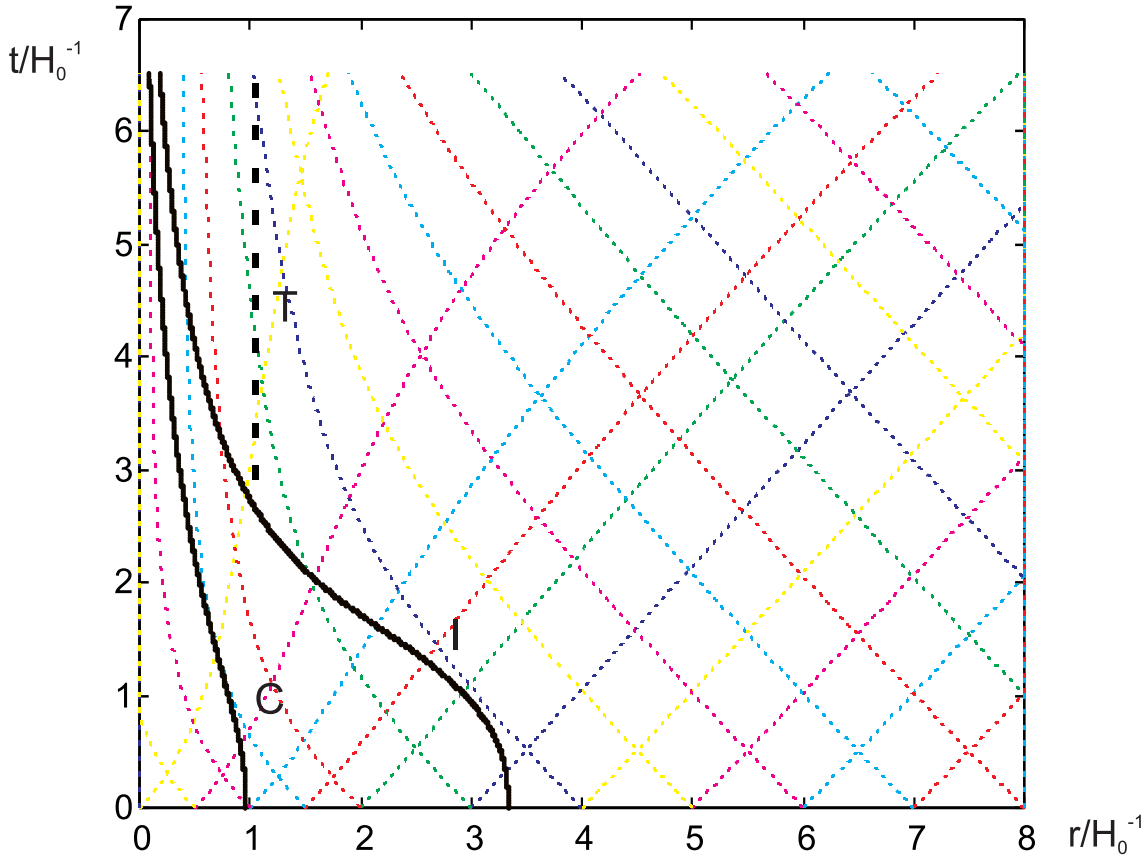}
\caption{
$r-t$ diagram illustrating the spacetime structure of an inflating monopole
of $\eta = 0.6\,m_p$.
The light dotted lines are ingoing and outgoing null geodesics.
The line $I$ is the boundary of the inflating region and
the line $C$ is the boundary of the monopole core where $\phi (t,r) =\eta /2$.
These two lines, $I$ and $C$, represent spacelike hypersurfaces.
The heavy dotted line $T$ is the location of the throat.
}
\label{fig=monconf}
\end{figure}

The overall spacetime structure of an inflating monopole is
illustrated in an $r-t$ diagram in Fig.\ref{fig=monconf}.  The inflating region of
spacetime is the region where the slow-roll condition is satisfied,

\begin{equation}
V'(\phi)\ll (48\pi G)^{1/2} V(\phi).
\label{eq=slowroll}
\end{equation}
The boundary of this region is shown by line $I$ in the diagram.  Also
shown is the boundary of the monopole core, which is defined as the
region inside the surface $\phi(t,r)=\eta/2$ (line $C$).  Both
surfaces are spacelike (as all surfaces of constant $\phi$ in the
slow-roll region).

\begin{figure}
\psfig{file=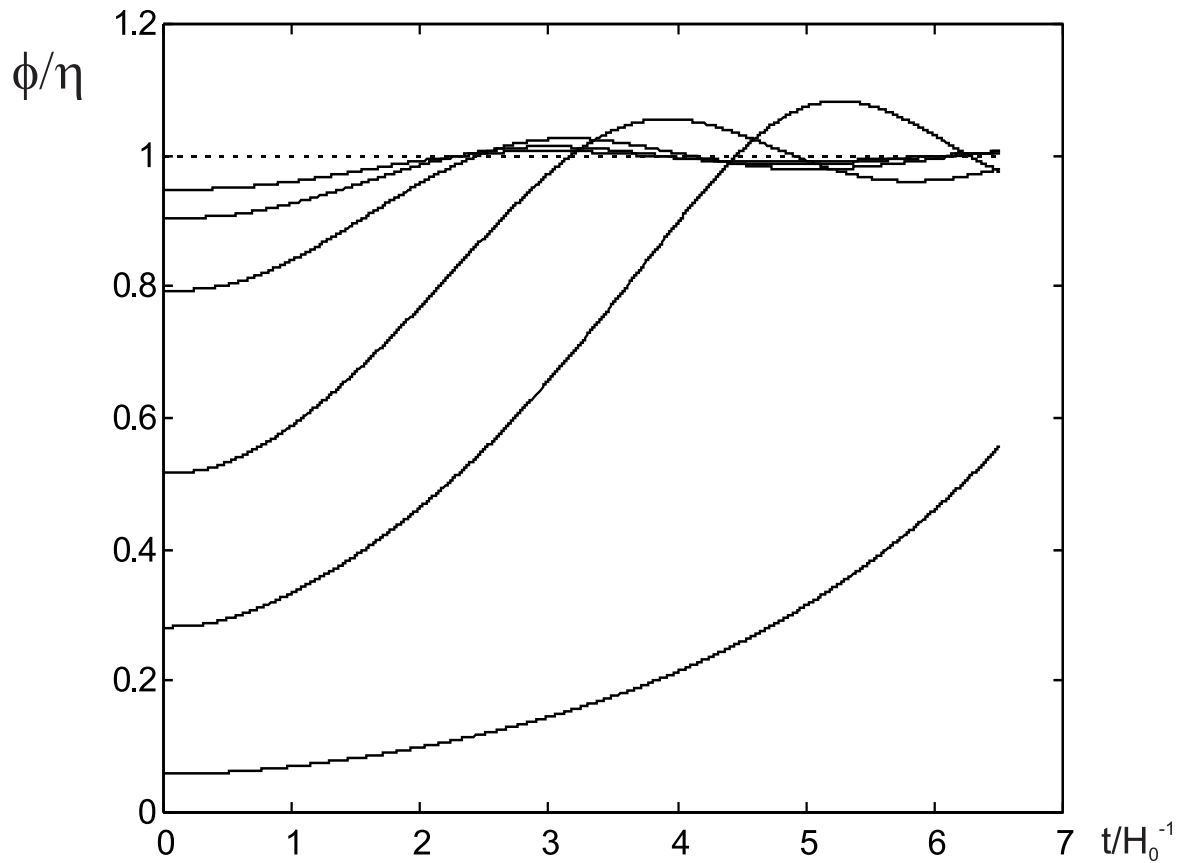}
\caption{
Evolution of the scalar field at a fixed $r$, for
$H_0r=0.1,0.5,1,2,3,4$ from the bottom.
The scalar field at small $r$ spends much time near the top of the potential
while it undergoes inflation. The field at large $r$ comes out of the 
inflating region earlier and oscillates about its vacuum expectation
value $\eta$.
}
\label{fig=montherm}
\end{figure}

Let us consider the evolution of the metric and the scalar field along
a timelike geodesic $r=const$ (the geodesic is not shown in the
figure).  In Fig.\ref{fig=montherm}, 
the field $\phi$ is shown as a function of $t$
for several values of $r$. We first choose $r<H_0^{-1}$, so that the
co-moving observer, whose trajectory this geodesic represents, 
starts inside the core at $t=0$.  As the field $\phi$ rolls down the
slope of the potential, the observer will come out of the core region,
and later out of the inflating region.  At this point $\phi$ will
start to oscillate about its vacuum expectation value $\eta$.  (This
oscillation is clearly visible in Fig.\ref{fig=montherm}.)  
The effective equation of
state for such an oscillating field, averaged over the period of
oscillation, is that of a pressureless dust.  Hence, our observer
emerges from an inflating to a matter-dominated region.  
In a more realistic model, the oscillations of $\phi$ would be damped
by particle production, resulting in a hot thermal radiation, but this
does not happen in our simple model.  

Observers with a co-moving coordinate $r$ between $H_0^{-1}$ and
$3H_0^{-1}$, start outside the monopole core, but otherwise follow the
same evolution and end up in a matter-dominated region.  On the other
hand, observers with $r>4H_0^{-1}$ have their starting points outside
the inflating region, and their surroundings are well described by the
asymptotic exterior metric (\ref{eq=asympmet}).  

Since the boundary of the inflating
region is spacelike, no observer can get into that region 
from the exterior or matter-dominated regions.  This can be seen by
examinimg the null geodesics, shown by dotted lines in Fig.\ref{fig=monconf}.

\begin{figure}
\psfig{file=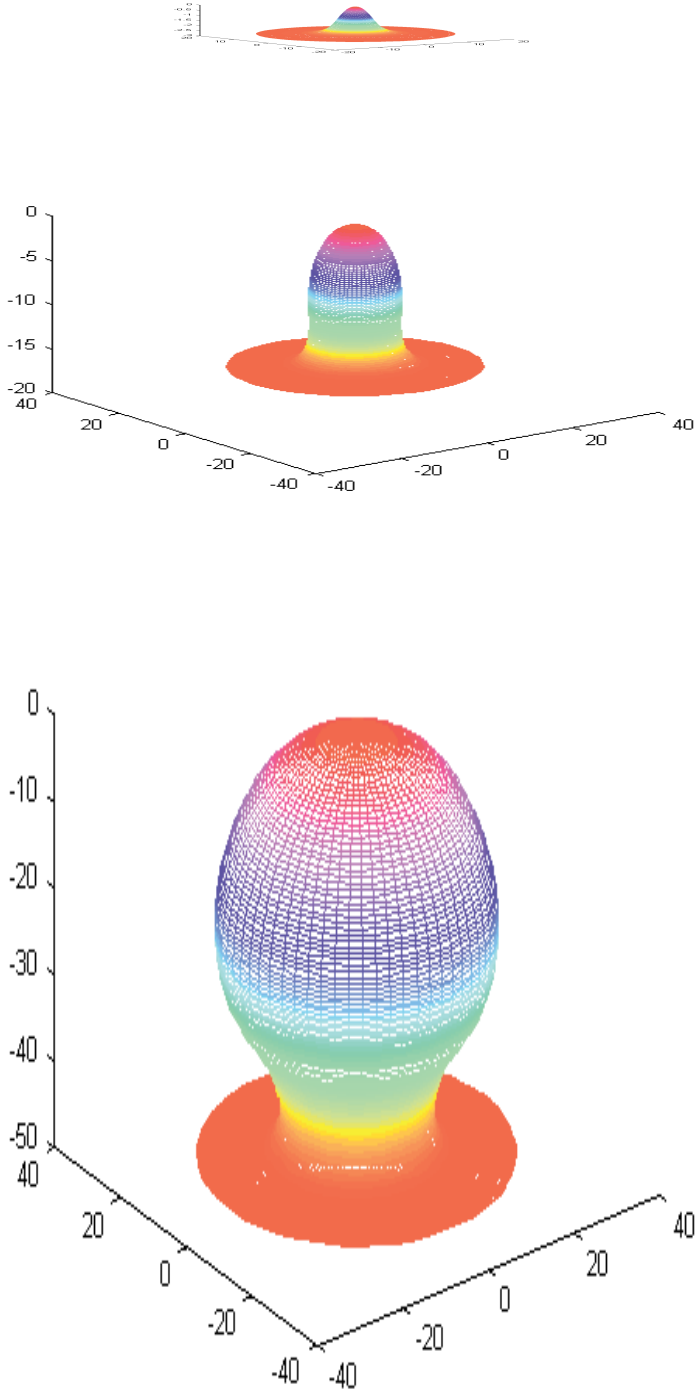}
\caption{
A 2-d slice of the inflating monopole ($\eta = 0.6\,m_p$) geometry 
embedded in a 3-d Euclidean space at $H_0t=1,3,5$ (from the top down).
The space gets curved around the monopole core and an inflating
balloon is formed later while the asymptotic region remains flat.
A throat is developed between these two regions.
}
\label{fig=moncyl}
\end{figure}

To illustrate the geometry of equal-time surfaces, $t=const$, in the
monopole spacetime, we shall use a lower-dimensional version of the
metric (\ref{eq=genmet}) with one of the angular dimensions suppressed.  
The 2-metric is then
\begin{equation}
ds^2=A(t,r)^2dr^2+B(t,r)^2r^2 d\theta^2.
\label{eq=spatialmet}
\end{equation}
Embeddings of this metric in a 3-dimensional Euclidean space 
\cite{cylplot} are shown in Fig.\ref{fig=moncyl} 
for several moments of time.  We see that the spatial
geometry is that of an inflating balloon connected by a throat to an
asymptotically-flat region at large $r$.  The throat is developed soon
after the onset of inflation.  Figure \ref{fig=moncyl} is similar to 
the figure in Sakai {\it et. al.} \cite{NS}, 
which shows a 3-d graph of proper radius $rB(t,r)$
vs. proper length $\int A(t,r)\, dr$ (rather than the embedded 2-geometry).

The location of the throat, determined from $d(Br)/dr=0$, 
is indicated by line $T$ in Fig.\ref{fig=monconf}.  
It is well approximated by a constant-$r$ line, $r \simeq H_0^{-1}$.

\section{Monopole evolution with cosmological initial conditions}

So far, we considered global monopoles coupled only to gravitation.
But cosmologically, monopoles are formed at a phase transition, and 
the monopole energy is taken from the energy of thermal radiation.  
In this section we shall consider the
evolution of a global monopole in a radiation background.

We use the same metric and scalar field ansatz as in section III.
The energy-momentum tensor of monopole-plus-radiation system is

\bq
	T_{\mu\nu}^{(tot)} = T_{\mu\nu}^{(m)} + T_{\mu\nu}^{(r)}\,,
\label{eq=emtensor}
\nq
The energy-momentum tensor of radiation is that of a perfect fluid with 
an equation of state $P = {1 \over 3}\rho$, 

\bq
	T_{\mu\nu}^{(r)} = (\rho + P) u_{\mu} u_{\nu} + P g_{\mu\nu}
 		= {\rho \over 3} (4 u_{\mu} u_{\nu} + g_{\mu\nu})\,,
\nq
where $u^{\mu}$ is the velocity 4-vector,

\[
	u^{\mu} = ({1 \over \sqrt{1-v^2}}, {v \over A \sqrt{1-v^2}}, 0, 0)\,.
\]

Prior to the phase transition, the space is filled with isotropic
radiation fluid, so the metric is that of FRW in the radiation-dominated
era, and

\bq
	\rho^{(r)} (t) = {3 \over 32\pi G t^2}\,.
\label{eq=rhot}
\nq
A realistic initial condition at the time of monopole formation, $t_0$,
is $\rho^{(tot)} (t_0,r) = \rho^{(m)} (t_0,r) + \rho^{(r)} (t_0,r) = 
const$, with $\rho^{(m)} \sim \rho^{(r)}$ near the monopole core.
For computational purposes, it is better to use a somewhat different
initial condition, 

\bq
	\rho^{(r)} (t_0,r) = const\,.
\label{eq=intl}
\nq
We tried them both and found the results to be very similar. The initial 
condition (\ref{eq=intl}) is preferable because it allows us to choose 
the initial value of $\rho^{(r)}$ somewhat lower than the core energy density
$V(0)$ and run the evolution until a later cosmic time. (We stop the run when
numerical instabilities develop in the inflating core region.)

The results presented below in this section were obtained using the following
initial conditions: $A(t_0,r)=B(t_0,r)=1$, $v(t_0,r)=0$, $\phi(t_0,r)$ 
the same as in the vacuum case, $\dot \phi (t_0,r) = 0$, 
$\rho^{(r)} (t_0,r) = 0.1\,V(0)$. The moment $t_0$ is determined by the 
initial value of $\rho^{(r)}$ from Eq. (\ref{eq=rhot}). 

\begin{figure}
\psfig{file=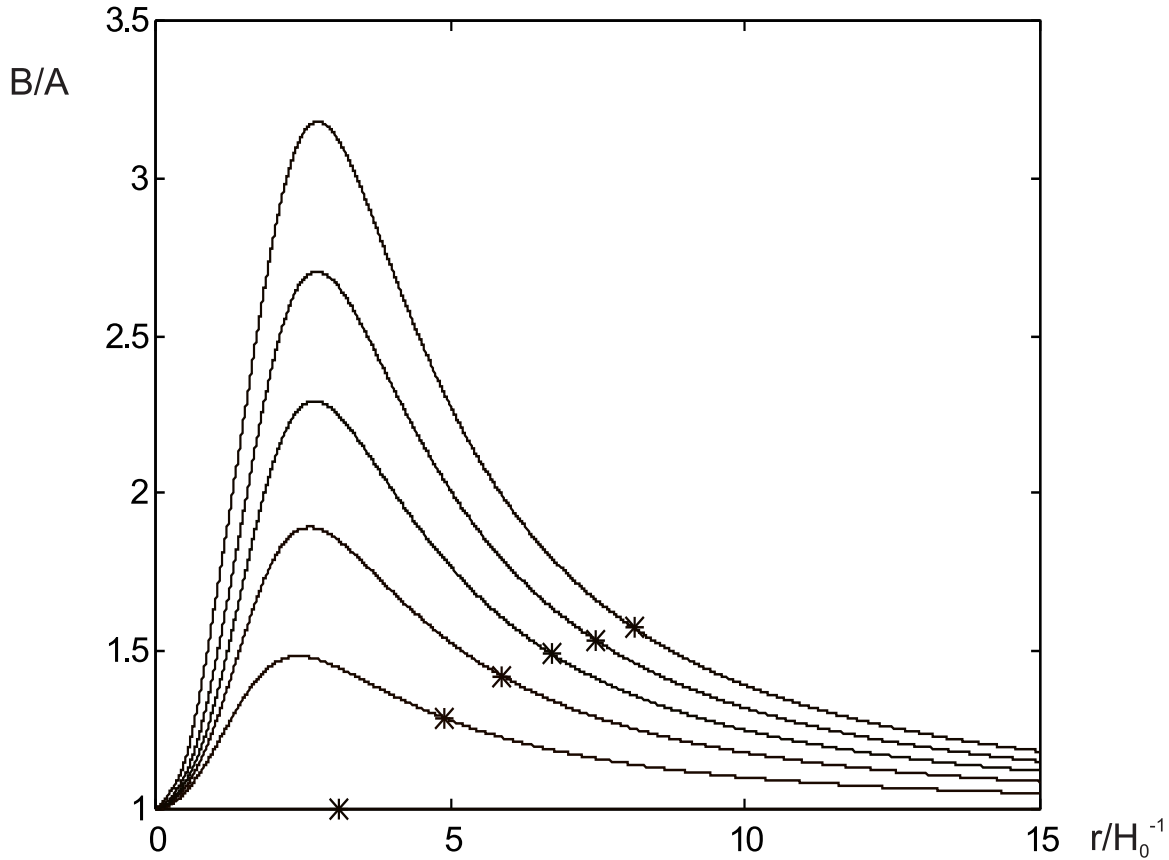}
\caption{
A plot of $B(t,r)/A(t,r)$ vs. $H_0r$ at $H_0(t-t_0)=0,1,2,3,4,5$
for a monopole of $\eta = 0.6\,m_p$ with the initial radiation density 
$\rho^{(r)} (t_0,r) = 0.1\,V(0)$.
Lower curves correspond to earlier times.
It shows that $A(t,r) \approx B(t,r)$ at small and large $r$.
The positions where $\rho^{(m)} = \rho^{(r)}$ are indicated by a star.
}
\label{fig=monrba}
\end{figure}

\begin{figure}
\psfig{file=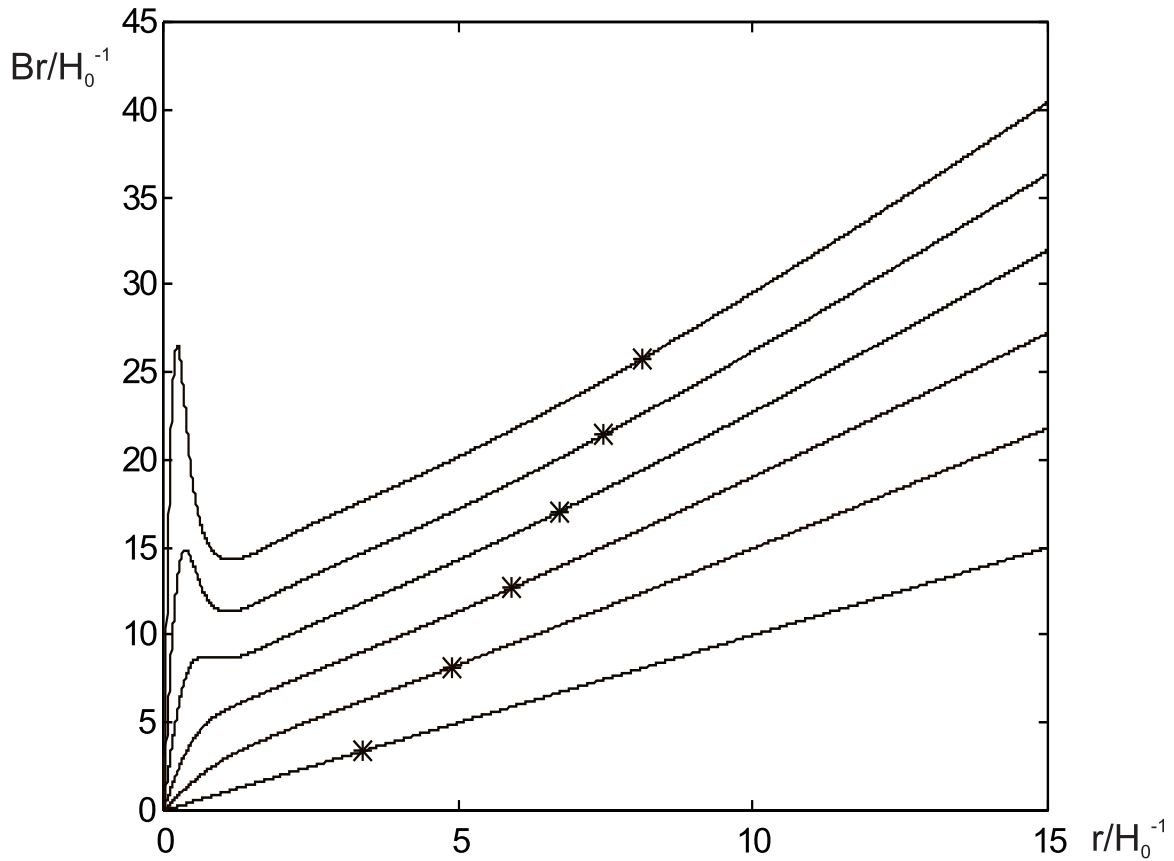}
\caption{
A plot of $H_0rB(t,r)$ vs. $H_0r$ at $H_0(t-t_0)=0,1,2,3,4,5$
for a monopole of $\eta = 0.6\,m_p$ with the initial radiation density
$\rho^{(r)} (t_0,r) = 0.1\,V(0)$.
Lower curves correspond to earlier times.
To the left of the stars, the plot is similar to that of
the vacuum solution in Fig.\ref{fig=monb}.
The positions where $\rho^{(m)} = \rho^{(r)}$ are indicated by a star.
}
\label{fig=monrb}
\end{figure}

\begin{figure}
\psfig{file=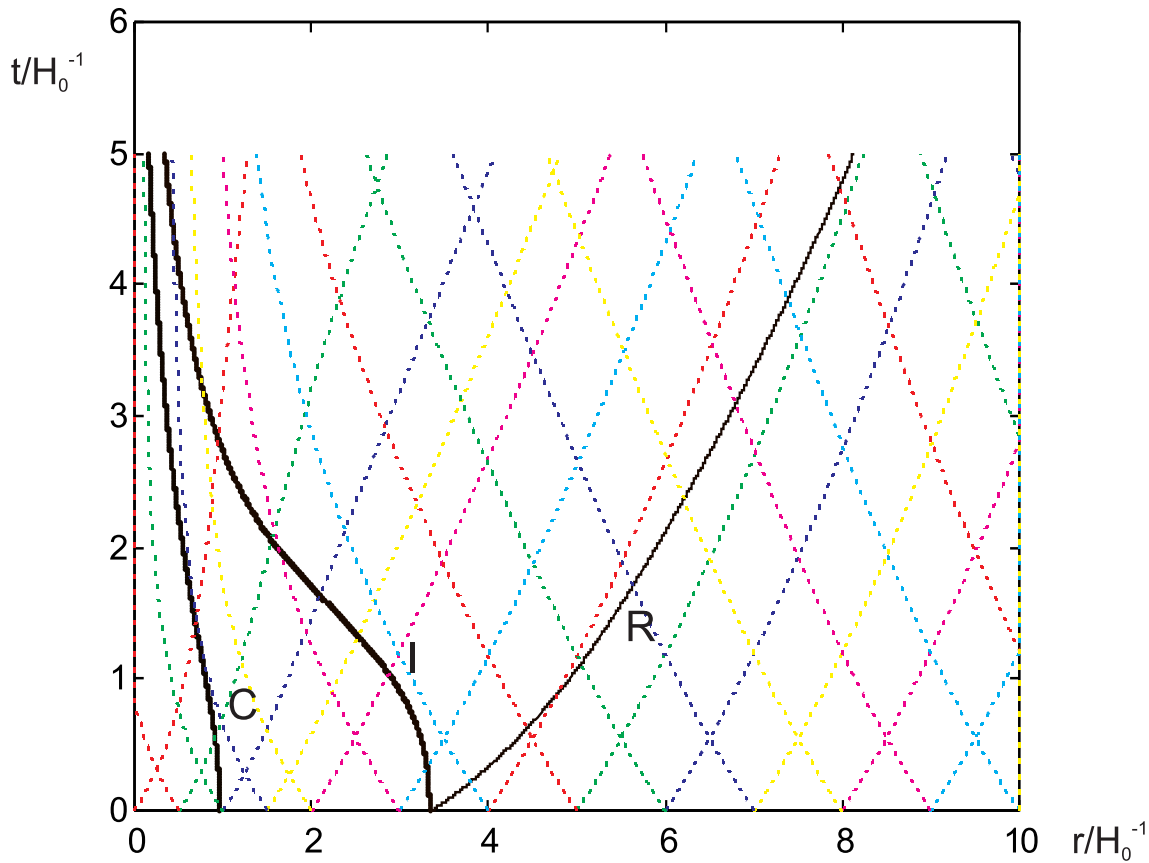}
\caption{
$r-t$ diagram illustrating the spacetime structure of an inflating monopole
of $\eta = 0.6\,m_p$ in a radiation background with $\rho^{(r)} (t_0,r)
= 0.1\,V(0)$.
The line $I$ is the boundary of the inflating region and
the line $C$ is the boundary of the monopole core where $\phi (t,r) =\eta /2$.
The line $R$ is the boundary where $\rho^{(m)} = \rho^{(r)}$.
}
\label{fig=monrconf}
\end{figure}

The field equations for the monopole-plus-radiation system, with a brief
outline of the method of their solutions, are given in Appendix.
Not surprisingly, we found that the critical symmetry breaking scale
$\eta_c$ is the same as in the vacuum case, $\eta_c \approx 0.33\,m_p$.
Our results for $\eta = 0.6\,m_p$ are presented 
in Figs. \ref{fig=monrba}-\ref{fig=monrconf}.

Fig.\ref{fig=monrba} shows that $B/A \approx 1$ in the inflating core 
and in the asymptotic region.
We have checked that the region of large $r$  exhibits the
usual radiation-dominated evolution, 

\begin{equation}
	A \approx B \approx (t/t_0)^{1/2}\,.
\label{eq=abt}
\end{equation}
In this region, the energy density of the monopole falls off as

\[
	\rho^{(m)} \sim {\eta^2 \over B(t,r)^2 r^2} 
		\sim {\eta^2 \over (t/t_0) r^2}\,.
\]
On the other hand, the radiation density is approximately homogeneous, 
but it damps more rapidly with time as in Eq. (\ref{eq=rhot}).
The two densities are comparable at

\begin{equation}
	r \approx (32\pi G\eta^2 t_0 / 3)^{1/2}\,\sqrt{t}
	  \approx 1.9\,(\eta / \eta_c)\,\sqrt{t_0t}\,.
\label{eq=eqdst}
\end{equation}
We can expect the region outside this surface to be well
approximated by an FRW radiation-dominated universe, and the interior
region to have spacetime structure similar to that of the vacuum
solution discussed in the preceeding section.  These expectations are
supported by our numerical results: we see a structure very similar
to Fig.\ref{fig=monb} emerging on the left hand side of Fig.\ref{fig=monrb}.

The overall spacetime structure is shown in Fig.\ref{fig=monrconf}, 
where the surface $\rho^{(m)} = \rho^{(r)}$ is indicated by line $R$. 
This line is reasonably well approximated by Eq. (\ref{eq=eqdst}).
(The fact that the line $R$ coincides with the boundary of the
inflating region $I$ at $t=t_0$ is a numerical coincidence.)

Comparing Eq. (\ref{eq=eqdst}) with the null geodesic in metric 
(\ref{eq=abt}), $r = 2\sqrt{t_0t}$, we see that the boundary 
(\ref{eq=eqdst}) expands faster than the speed of light 
(unless perhaps when $\eta$ is very close to $\eta_c$). 
Hence, physical observers can not get from the exterior vacuum region 
(dominated by the scalar field) to the radiation-dominated region.

\section{Conclusions}

We have investigated the spacetime structure of an inflating global
monopole, both inside and outside the inflating core. 
We found that the inflating region near the core is surrounded by a
``matter'' region, where the dominant contribution to the energy is
that of an oscillating scalar field. The inflating region and most
of the matter region are contained in an expanding ``baloon''
which is connected by a thoroat to the exterior region.

The exterior metric at large distances from the throat is well
approximated by Eq. (\ref{eq=asympmet}). This metric is obtained by
analytic continuation of the static monopole metric 
(\ref{eq=manuelmet}), followed by a coordinate transformation
(\ref{eq=cdtransf}).
It describes a non-stationary spacetime with a highly anisotropic
expansion.

A spaceship from the exterior region can pass through the throat 
and get into the matter-dominated region, but it cannot get into 
the inflating region. The inflating region is bounded by a spacelike
hypersurface and is seen by observers in the matter region as an epoch
in their past. 

To simulate the formation of an inflating monopole at a cosmological
phase transition, we studied the monopole evolution in a radiation 
background. At the initial moment, the radiation energy dominates
everywhere, except in the central region near the core.
However, this energy is red-shifted faster than that of the scalar
field, and as a result the boundary of the radiation-dominated region
is pushed to larger and larger distances. The spacetime structure
emerging at large $t$ inside this boundary is very similar to that
in the vacuum case without radiation.

It would be interesting to perform a similar analysis for topological
defects other than global monopoles. In the case of gauge monopoles,
we expect \cite{AV} the exterior region to be described by the
Reissner-Nordstr$\o$m metric. For strings and domain walls, the situation 
is not so clear. This problem is now being investigated.

\acknowledgements

We are grateful to Nobuyuki Sakai for helpful correspondence and to
the National Science Foundation for partial support.

\appendix
\section*{Field equations for the monopole-radiation system}

Einstein's equations with the metric (\ref{eq=genmet}) and energy-momentum 
tensor (\ref{eq=emtensor}) are:

\bqr
	- G_0^0 &=&
      	K_2^2 (2K-3K_2^2) - 2{B'' \over A^2B} -{B'^2 \over A^2B^2}
	+2{A'B' \over A^3B} -6{B' \over A^2Br} +2{A' \over A^3r}
	-{1 \over A^2r^2} + {1 \over B^2r^2}  \nonumber\\
	&=& 8 \pi G \l[ \,{\dot{\phi}^2 \over 2} + {\phi'^2 \over 2A^2}
	+{\phi^2 \over B^2r^2} + {\lambda \over 4}(\phi^2 - \eta^2)^2
	+{1 \over 3} \l( {4 \over 1-v^2} -1 \r) \rho\, \r] \,,
\label{eq=energy}
\nqr
\bq
	{1 \over 2} G_{01} = 
	K_2^{2\prime}  + \l( {B' \over B} + {1 \over r} \r) \l( 3K_2^2 - K \r) = 
	4\pi G \l( \dot{\phi} \phi' -{4 \over 3}\,{v \over 1-v^2}\,A\,\rho \r) \,,
\label{eq=K2}
\nq
\bqr
	{1 \over 2}\,(G_1^1 + G_2^2 + G_3^3 - G_0^0) &=&
      	\dot{K} - (K_1^1)^2 - 2(K_2^2)^2 \nonumber\\
	&=& 8\pi G \l[ \, \dot{\phi}^2 - {\lambda \over 4}(\phi^2 - \eta^2)^2
	+ {1 \over 3} \l( 1 + 2\, {1+v^2 \over 1-v^2} \r) \rho\, \r] \,,
\label{eq=K}
\nqr
where, 
\bq
	K_1^1 = - {\dot{A} \over A}\,, \quad
	K_2^2 = K_3^3 = - {\dot{B} \over B}\,, \quad
	K = K_i^i\,.
\label{eq=Ks}
\nq
The conservation of energy-momentum tensor, 
$\, T^{(r)\,\mu\nu}_{\qquad ;\nu} = 0\,$,
gives two equations for $\rho$ and $v$:

\bq
      	(3 - v^2)\, \dot{v} = -{2v \over A}\, v' 
	-{3(1-v^2)^2 \over 4A}\, {\rho' \over \rho}
	 + 2(K_1^1 -K_2^2)\,v\,(1-v^2)
	+ 2 \l( {B' \over B} + {1 \over r} \r) {v^2(1-v^2) \over A} \,,
\label{eq=v}
\nq
\bq
      	{\dot{\rho} \over \rho} = {8v \over 3(1-v^2)^2}\, \dot{v}
	- {4(1-3v^2) \over 3A(1-v^2)^2} \,v' +K_1^1 \,{4(1-3v^2) \over 3(1-v^2)}
	+ K_2^2 \,{8 \over 3(1-v^2)} 
	- \l( {B' \over B} + {1 \over r} \r) {8v \over 3A(1-v^2)}\,. 
\label{eq=rho}
\nq

The field equation for $\phi$ is

\bq
	\ddot{\phi} - K\,\dot{\phi} - {\phi'' \over A^2}
	- \l( -{A' \over A} + {2B' \over B} + {2 \over r} \r)\,
	{\phi' \over A^2} + {2\phi \over B^2 r^2} + {dV \over d\phi} = 0.
\label{eq=field}
\nq

With the initial conditions given in section V,
$K(t_0,r)$ and $K_2^2(t_0,r)$ are evaluated by the Hamiltonian and the momentum
constraint equations (\ref{eq=energy}), (\ref{eq=K2}).
In the next time step, $v(t,r)$ and $\rho (t,r)$ are calculated by 
(\ref{eq=v}) and (\ref{eq=rho}). 
$A(t,r)$ and $B(t,r)$ are calculated by (\ref{eq=Ks}), $\phi (t,r)$ 
by the scalar field equation (\ref{eq=field}) and $K_2^2$, $K$ 
by the constraint equations (\ref{eq=K2}), (\ref{eq=K}).
We apply the regularity condition $K_1^1(t,r=0) = K_2^2(t,r=0)$ at the origin.

The field equations in the vacuum case are obtained by setting $\rho = 0$
in Eqs. (\ref{eq=energy})-(\ref{eq=K}).


\begin{references}
\bibitem{cylplot}
The metric of a 3-d Euclidean space is
$ds^2 = dz^2 + d\rho^2 + \rho^2 d\theta^2$. Comparing this with
(\ref{eq=spatialmet}), we have $\rho = Br$ and
$z' = \pm \l[ A^2 - (B'r+B)^2 \r]^{1/2}$, where the primes indicate
derivatives with respect to $r$. The embedding is specified by the
functions $\rho (r)$ and $z(r)$ which can be found from the above
equations. The vertical axis in Fig.\ref{fig=moncyl} is the $z$-axis.
\end{references}
\end{document}